\documentclass[pdflatex,sn-nature]{sn-jnl}

\usepackage{amsmath}  % needed for \tfrac, \bmatrix, etc.
\usepackage{amssymb}
\usepackage{amsfonts} % needed for bold Greek, Fraktur, and blackboard bold
\usepackage{graphicx} % needed for figures
\usepackage{physics}  % needed for vectors
\usepackage{color}    % needed for colors and color comments
\usepackage{ulem}     % needed for strikethrough

\definecolor{DCW}{rgb}{1,0,0}
\definecolor{DH}{rgb}{1,0.6,0}
\definecolor{ML}{rgb}{0,0.8,0.2}

\begin{document}

% Be sure to use the \title, \author, \affiliation, and \abstract macros
% to format your title page.  Don't use lower-level macros to  manually
% adjust the fonts and centering.

\title{The hardest-hit home run?}
% In a long title you can use \\ to force a line break at a certain location.

%When submitting the manuscript for review, do not include the author's name or institution
%\author{Daniel V. Schroeder}
%\email{dschroeder@weber.edu} % optional
%\altaffiliation[permanent address: ]{101 Main Street, Anytown, USA} % optional second address
% If there were a second author at the same address, we would put another 
% \author{} statement here.  Don't combine multiple authors in a single
% \author statement.
%\affiliation{Department of Physics, Weber State University, Ogden, UT 84408-2508}
% Please provide a full mailing address here.

\author*[1,2,3]{\fnm{Donald C.} \sur{Warren} \sfx{III}}\email{dwarren@fit.edu}
\affil*[1]{\orgdiv{Department of Aerospace, Physics, \& Space Sciences}, \orgname{Florida Institute of Technology}, \orgaddress{\street{150 W University Blvd}, \city{Melbourne}, \postcode{32901}, \state{FL}, \country{USA}}}
\affil[2]{\orgdiv{Astrophysical Big Bang Laboratory}, \orgname{RIKEN}, \country{Japan}}
\affil[3]{\orgdiv{iTHEMS Program}, \orgname{RIKEN}, \country{Japan}}

% See the REVTeX documentation for more examples of author and affiliation lists.

\date{\today}

\abstract{
We present a problem to be assigned or done as an in-class activity in an upper-division undergraduate course on computational physics.  The problem involves a home run hit by Mickey Mantle on May 22, 1963, which he famously called ``the hardest ball I ever hit''.  Is this home run truly one for the record books, or has it been eclipsed by players in the modern era?  Modeling the trajectory of a baseball involves consideration of both wind resistance and the Magnus effect, and is an interesting application of numerical solution of ordinary differential equations.  Ultimately, the answer is that Mantle would compare favorably to the most powerful batters currently playing, but to arrive at that conclusion we must reflect on the plausibility of results and sources of uncertainty.
}
% AJP requires an abstract for all regular article submissions.
% Abstracts are optional for submissions to the "Notes and Discussions" section.

\maketitle % title page is now complete

\section{Introduction} % Section titles are automatically converted to all-caps.
% Section numbering is automatic.

Baseball is a statistician's dream: a sport with well over a century of records kept to an overwhelming level of detail.  With so much history, so many famous players, and so many games played, even the most exciting events have happened many times. There have been nearly 240,000 games played over almost 150 years;\cite{bbref_totals} 335,381 home runs have been hit since  the 1876 season started;\cite{bbref_totals} and even so-called ``perfect games'' have happened two dozen times.\cite{mlb_perfectgames}  But amongst all of this, when Mickey Mantle (1931-1995), one of the sport's most famously powerful hitters, explicitly called out one of his home runs as ``the hardest ball I ever hit,''\cite{NYTcoverage} his claim merits particular attention.  

\begin{figure}[h!]
% The bracketed code determines the figure's placement:  "h" stands for 
% "here", telling LaTeX to put the figure as close to the current location 
% as possible.  The ! overrides LaTeX's tendency to try to find a location 
% that it thinks is better.  But don't agonize over the exact figure placement 
% in your submitted manuscript.  For your initial submission, just make sure 
% each figure is reasonably close to where it's first referenced.
\centering
\includegraphics[width=0.6\textwidth]{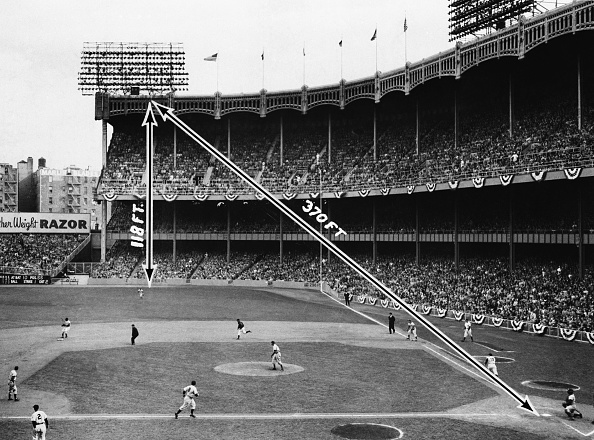}
\caption{A picture of the original Yankee Stadium, showing the distance a baseball would have to travel to hit the facade over right field, as Mantle's home run did.  (This photo is illustrative only.  The actual game on May 22, 1963 took place at night, and we know of no surviving photographs.)  \textit{Image credit}: Bettmann, via Getty Images.
}
\label{fig:homerun_photo}
\end{figure}

Starting from home plate and an estimated three feet off the ground, the baseball very nearly cleared the decorative facade of Yankee Stadium almost 118 feet above ground level, traveling a horizontal distance of 351 feet before striking the facade.  The ball almost became the first home run to be hit out of Yankee Stadium, missing this feat by mere inches.  Perhaps even more amazingly, observers agree that the baseball was \textit{still rising} when it hit the facade, thus not having reached the highest point in its trajectory.

Attendees claimed that the sound of the bat hitting the ball was like a cannon firing, but, even allowing for exaggeration, it must be assumed that the ball left the bat with a truly tremendous speed.  But just how tremendous, and has Mantle been overshadowed by modern players?  Comparison across eras of baseball is an inherently fraught endeavor.  The rules have changed over the years, athletes are more specialized and more highly trained than ever before, and even local microclimates associated with different venues can impact the outcomes of games. Coors Field in Denver, for example, produces longer hits than other fields because of the reduced air density at that elevation. While an analysis of Mantle's home run is not without significant uncertainties, it makes for a fascinating case study of projectile physics. 

In the activity described in this paper, students will numerically solve the ordinary differential equations (ODEs) associated with baseball flight, making some assumptions along the way, to solve a boundary-value problem and determine the initial conditions of Mantle's home run.  This is possible because one thing that has \textit{not} changed since the start of professional baseball is physics. We can, in fact, estimate just how hard Mantle really hit that ball, and compare it to power hitters in the modern era (since 2015) where Statcast\cite{mlb_statcast} has been providing high-precision tracking of baseball trajectories.  Answering this question will require dealing with substantial sources of uncertainty, and we will not, in the end, arrive at a single answer that both meets known conditions and is physically plausible.  But there is no harm in saying ``We don't know'' in a classroom activity like this one.

\section{Theory of baseball flight}
\label{sec:theory}

An extraordinary amount of effort has been spent on determining the theory of how baseballs move once thrown or batted; see, for example, Refs.\ \cite{adair2002, nathansite, Cross2012, EscaleraSantos_etal_2019} and the numerous references therein.  In the absence of air resistance, the trajectory of a baseball is treatable using the equations of introductory kinematics, making the problem straightforward if too simple for a computational physics course.  Cast as second-order ordinary differential equations, the following formulae govern the motion of projectiles:
\begin{subequations}
\label{eq:proj}
\begin{align}
    a_{x,\mathrm{proj}} &= \frac{ d^{2} x }{ dt^{2} } = 0 \\
    a_{y,\mathrm{proj}} &= \frac{ d^{2} y }{ dt^{2} } = -g
\end{align}
\end{subequations}
These expressions assume  that the motion is restricted to two dimensions and that the ball stays close enough to Earth's surface that the gravitational acceleration is effectively constant;  for our purposes the traditional value of $g = 9.807$ m/s$^{2}$ will suffice given other uncertainties we will encounter.

However, baseball games are not played in a vacuum, so we must account for the effects air has on the flight of the ball.  For the purposes of this activity, that means considering two key effects: air resistance and the Magnus effect.

The goal here is not a thorough, first-principles derivation of these complications; many such derivations and explanations already exist in the literature.\cite{nathansite, Cross2012, EscaleraSantos_etal_2019}  Rather, we briefly summarize the key features and how they would be implemented in an undergraduate course on numerical methods.

\subsection{Wind resistance}

As a baseball travels through air, it must push air out of the way.  By Newton's Third Law, the air exerts a force on the baseball that opposes the motion of the ball.  The acceleration due to air resistance can be approximated as quadratic in the speed.  Written in components, it takes the form
\begin{subequations}
\label{eq:windres}
\begin{align}
    a_{x,D} &= -\frac{1}{2 m} \rho_\mathrm{air} c_{d} A v \, v_{x} \\
    a_{y,D} &= -\frac{1}{2 m} \rho_\mathrm{air} c_{d} A v \, v_{y}, 
\end{align}
\end{subequations}
where $\rho_\mathrm{air}$ is the density of the air, $A = \pi r^{2}$ is the cross-sectional area of the ball, $v$ the instantaneous speed of the ball ($v = \sqrt{v_{x}^{2} + v_{y}^{2}}$), and $c_{d}$ is an empirically-fitted drag coefficient that relates the cross-sectional area of the ball to the actual amount of drag force experienced.  The minus signs and the components $v_{x}$ and $v_{y}$ of the velocity show that the acceleration is directed opposite to the motion of the ball. There is evidence that the drag coefficient also depends on the spin of the ball,\cite{lyu_etal_2022} but in the interest of simplicity we neglect this comparatively minor effect.

A further wrinkle is the wind itself.  The velocity $\va{v}$ of the ball in Equation~\ref{eq:windres} is more properly $\va{v}_\mathrm{ball} - \va{v}_\mathrm{wind}$. A ball traveling precisely with the wind does not experience any wind resistance at all, while wind blowing against the ball increases the rate at which it  must displace air molecules.  We return to this point in Section~\ref{sec:conditions} when we discuss the weather at Yankee Stadium on May 22, 1963.

The drag coefficient $c_{d}$ in Equation~\ref{eq:windres} is determined empirically.  Air is not an ideal zero-viscosity fluid; the drag coefficient depends on the speed with which the ball travels through the air.  The following equation is a fitted formula that is accurate to within a few percent for all reasonable ball speeds (see Ref \cite{nathan_trajectory}, citing Ref \cite{adair2002}):
\begin{equation}
    %c_{d} = 0.29 + \frac{ 0.22 }{ 1 + e^{(v - 32.37)/5.2} }
    c_{d} = 0.5 - \frac{ 0.227 }{ 1 - e^{(33.14-v)/6.40}},
\end{equation}
where $v$ is the speed of the ball in meters per second.

\subsection{The Magnus effect}

The Magnus effect is a property of spinning shapes traveling through the air. When the center of mass of a baseball travels at some speed $\va{v}$, the sides have different speeds with respect to the wind due to the rotation of the ball.  This difference affects how the air flows around and in the wake of the ball, leading to a net force.\footnote{This effect has been known since at latest Isaac Newton's time, and was studied and codified by its namesake Heinrich Magnus.\cite{wikiMagnus} In baseball, the Magnus effect was applied to pitching curveballs in the mid 19th century; it was ``disproven'' in the mid 20th century,\cite{LIFEnocurveballs} and reproven shortly thereafter;\cite{Verwiebe1942, Sutton1942} and it has since seen sustained effort to conclusively understand the physics and to quantify its impact on the sport of baseball.\cite{Briggs1959, Rex1985, WattsFerrer1987, Nathan2008}}  A more thorough description of how baseball spin modifies trajectories can be found in Ref.\ \cite{Nathan2008}.  Ultimately, however, this effect still appears to have no first-principles theoretical explanation; our understanding is based on exhaustive experimentation.

The computation of this force is more involved than that of wind resistance: the force depends not only on the velocity $\va{v}$ of the ball (which technically accounts for any wind) and the angular velocity $\va{\omega}$, but on the relationship between those two vectors.\cite{EscaleraSantos_etal_2019}
\begin{equation}
    \va{F}_{M} = \frac{1}{2} \rho_\mathrm{air} c_{M} A v \,\, \frac{ \va{\omega} \times \va{v} }{ \omega }.
\end{equation}
Once again, the magnitude of the force depends on the air density, the cross-sectional area of the ball, the speed of the ball, and on another empirically-determined coefficient, $c_{M}$.  The value of $c_{M}$ depends on the dimensionless spin factor $S = r\omega/v$ of the baseball (not to be confused with the state vector $\va{S}$ in Section~\ref{sec:num}); this is the ratio of the maximum linear speed of the spinning ball's surface, $r \omega$  (with $\omega$ in radians per second) to the ball's linear speed $v$ through the air.  With $S$ calculated, $c_{M}$ is found via\cite{nathan_trajectory}
\begin{equation}
    c_{M} = \frac{ 1.12 S }{ 0.583 + 2.333 S }.
    \label{eq:cM}
\end{equation}

Because the cross-product will point perpendicular to $\va{v}$, the only way for our system to remain two-dimensional is if $\va{\omega}$ points solely in the $z$ direction: either perfect topspin or perfect backspin.  We will assume for the moment that the baseball has only backspin, not only because it simplifies the equations but also because it results in the lowest possible initial speed for a batted ball (more on this in Section~\ref{sec:plausibility}).  With this assumption, we can compute the cross product and write down the resulting acceleration due to the Magnus effect for use alongside Equations~\ref{eq:proj}--\ref{eq:windres},
\begin{subequations}
\label{eq:magnus}
\begin{align}
    a_{x,M} &= - \frac{1}{2m} \rho_\mathrm{air} c_{M} A v \cdot v_{y} \\
    a_{y,M} &= \phantom{-} \frac{1}{2m} \rho_\mathrm{air} c_{M} A v \cdot v_{x}.
\end{align}
\end{subequations}
Note that the $y$-component of this the acceleration is positive.  The Magnus effect, when the baseball has the appropriate spin, serves to lift the ball further into the air and can at least partially counteract the effects of gravity and wind resistance.

\section{Numerical setup}
\label{sec:num}

This activity is intended for a class on computational methods, so in this section we combine the theory into a single differential equation and outline a procedure for solving it.  For a fuller explanation of the process and nuances of solving differential equations, see, among many others, Refs.\ \cite{newman,gezerlis}.  Numerous codes in a variety of languages exist for this purpose, including the exceptional Trajectory Calculator in Ref.\ \cite{nathan_trajectory}.  As the goal here is to guide students through applying differential equations, however, we suggest producing a solver from scratch rather than relying on fully-complete products; a Python code is nonetheless provided in the supplementary material.\cite{pythoncode}

The state vector of the baseball while it is in flight is specified by the combination its position and velocity vectors,
\begin{equation}
    \va{S} = \{ \va{r}, \va{v} \} = \{ x, y, v_{x}, v_{y} \}.
\end{equation}
Extending this to three dimensions is not overly challenging and might make for a good student project.

We wish to solve an ODE that relates the current state vector to its rate of change,
\begin{equation}
    \dot{\va{S}} = \va{f}(\va{S}),
    \label{eq:basic_ODE}
\end{equation}
where $\va{f}$ is a vector function with one component for each component of $\dot{\va{S}}$. (Note that $\va{f}$ might sometimes explicitly depend on time in addition to the state vector, i.e. $\va{f}(\va{S},t)$, if for example there were a time-dependent force. That is not the case in this exercise, so we use the simpler notation.)

Equation~\ref{eq:basic_ODE} can be related  to the equations presented in Section~\ref{sec:theory} via the definitions of velocity and acceleration,
\begin{equation}
    \dot{\va{S}} = \{ \dot{x}, \dot{y}, \dot{v}_{x}, \dot{v}_{y} \} = \{ v_{x}, v_{y}, a_{x,\mathrm{tot}}, a_{y,\mathrm{tot}} \} = \va{f}(\va{S}),
\end{equation}
in which $a_{x,\mathrm{tot}}$ and $a_{y,\mathrm{tot}}$ include all relevant accelerations; the last equality is repeated to emphasize the relationship between  physical quantities and their numerical representations.  Incorporating the results of Section~\ref{sec:theory} leads to the following four expressions:
\begin{subequations}
\label{eq:derivs}
\begin{align}
    \dot{x} &= v_{x} \\
    \dot{y} &= v_{y} \\
    \dot{v_{x}} &= a_{x,\mathrm{proj}} + a_{x,D} + a_{x,M} = \phantom{-g} -\frac{1}{2 m} \rho_\mathrm{air} c_{d} A v \, v_{x} - \frac{1}{2m} \rho_\mathrm{air} c_{M} A v \, v_{y} \\
    \dot{v_{y}} &= a_{y,\mathrm{proj}} + a_{y,D} + a_{y,M} = -g - \frac{1}{2 m} \rho_\mathrm{air} c_{d} A v \, v_{y} + \frac{1}{2m} \rho_\mathrm{air} c_{M} A v \, v_{x}.
\end{align}
\end{subequations}
These equations represent $\va{f}(\va{S})$ in Equation~\ref{eq:basic_ODE}.

Numerical solution of ODEs must be done approximately.  The simplest (and therefore, the crudest) approximation is Euler's method, which treats the derivative $\dot{\va{S}}$ as a single step in time with constant slope, leading to
\begin{align}
    \va{f}(\va{S}) &= \dot{\va{S}} \approx \frac{ \va{S}_{f} - \va{S}_{i} }{ \Delta t} \nonumber \\
    &\implies \va{S}_{f} = \va{S}_{i} + \Delta t \,\, \va{f}(\va{S})
\end{align}
A much better approximation which allows for fewer/larger time steps for the same level of accuracy can be achieved by taking small intermediate steps within the larger step of $\Delta t$ and by cleverly combining the results to eliminate error terms.  An example of this is the 4$^\mathrm{th}$-order Runge--Kutta formula, which can be expressed as
\begin{subequations}
\label{eq:RK4}
\begin{align}
    \va{k}_{1} &= \Delta t \,\, \va{f}(\va{S}) \\
    \va{k}_{2} &= \Delta t \,\, \va{f}(\va{S} + \frac{1}{2} \va{k}_{1}) \\
    \va{k}_{3} &= \Delta t \,\, \va{f}(\va{S} + \frac{1}{2} \va{k}_{2}) \\
    \va{k}_{4} &= \Delta t \,\, \va{f}(\va{S} + \va{k}_{3}) \\
    \va{S}_{f} &= \va{S}_{i} + \frac{1}{6}\left( \va{k}_{1} + 2 \va{k}_{2} + 2 \va{k}_{3} + \va{k}_{4} \right) \label{eq:y_accel}
\end{align}
\end{subequations}
The computations presented later use the above formula, though it is up to the discretion of the instructor which approach will be asked of the students.

It is important at this point to note that both the Euler method and Runge--Kutta methods (of any order) take information about the initial conditions and propagate it forward in time to reach the final time of interest.  For the problem of evaluating Mantle's home run, we do not have all of the initial conditions.  Instead, we have a mixture of initial and final conditions, and so we are left with a Boundary Value Problem (BVP) rather than an Initial Value Problem (IVP).

To solve BVPs, we must make assumptions about the unknown initial conditions (velocity, in the present case), solve the ODE as if it were a regular IVP, and at the end compare the output to the known final conditions (position).  If the agreement is close enough, then we have found \textit{a} solution to the BVP, but not necessarily the unique solution.  The process can be automated as needed and depending on the sophistication of the students, but simple ``guess and check'' will often work well enough for a small number of cases, if more slowly than the automated algorithm provided in the supplementary material.\cite{pythoncode}

\section{Boundary, Weather, and Other Conditions}
\label{sec:conditions}

Having discussed both the equations we wish to solve and the process by which we will solve them, one issue remains: providing numerical values for all of the quantities from Section~\ref{sec:theory}.

We begin with the values that make this home run a BVP, and therefore  more challenging than a standard IVP:
\begin{align}
    x(0) &= 0 \nonumber \\
    y(0) &= 0.9~\mathrm{m}~(3~\mathrm{ft}) \nonumber \\
    x(t) &= 106.9~\mathrm{m}~(351~\mathrm{ft}) \nonumber \\
    y(t) &= 36~\mathrm{m}~(118~\mathrm{ft})
\end{align}
Not only do we have a mixture of initial and final values, but we also don't know \textit{when} to stop tracking the flight of the ball, only \textit{where}.  If we wish to turn this into an IVP, we must make assumptions about the remaining two initial conditions, $v_{x}(0)$ and $v_{y}(0)$.  We are interested in the speed of the ball as it leaves the bat, so we recast those conditions in terms of two different unknown quantities, the initial speed $v_{0}$ and the launch angle $\theta_{0}$ of the ball:
\begin{align}
    v_{x}(0) &= v_{0} \cos \theta_{0} \nonumber \\
    v_{y}(0) &= v_{0} \sin \theta_{0}.
\end{align}
Launch angles of less than $0^{\circ}$ would be hit toward the ground, launch angles of $90^{\circ}$ would be hit straight into the air, and launch angles of $25-30^{\circ}$ are associated with a high likelihood of home runs.\footnote{Compare the expectation of $45^{\circ}$ to maximize a projectile's range in a vacuum.}

We wish to obtain an estimate for the initial speed $v_{0}$, meaning all other quantities must be known: the mass $m$ and cross-sectional area $A$ of the baseball, the density $\rho_\mathrm{air}$ of the air in Yankee stadium on 22 May 1963, the launch angle $\theta_{0}$, and the spin rate $\omega$ of the ball.
\begin{itemize}
    \item The mass and area of the baseball are fixed by the rules of professional baseball.  Its mass is approximately $m = 0.145$~kg.  Its radius is 37~mm, and hence its cross-sectional area is $A = 0.0043$~m$^{2}$.
    \item The density of the air depends on its temperature, pressure, and relative humidity.  Historical weather records for  New York are available online,\cite{weatherunderground} and show that when the night game might plausibly have finished, the ambient air temperature was $T = 18^{\circ}$C (65$^{\circ}$F) with a relative humidity  $RH = 39$\% and an atmospheric pressure $P = 101.65$~kPa (30.02 inches of mercury).  While calculators exist online,\cite{airdensity_calc} these quantities may instead be plugged into the following formula,\cite{airdensity_eqs}
    \begin{equation}
        \rho_\mathrm{air} = 3.4837 \cdot \frac{ P }{ T + 273.15 } - 0.0080434 \cdot \frac{ RH }{ T + 273.15 } \cdot \mathrm{exp}\left[ \frac{17.27 T}{T + 237.3} \right]
    \end{equation}
    where $P$ is measured in kPa, $T$ in $^{\circ}$C, and $RH$ in percent, with $\rho_\mathrm{air}$ in kg/m$^{3}$.  Either approach yields an air density $\rho_\mathrm{air} = 1.21$~kg/m$^{3}$.
   
    \item Wind was fairly significant on the evening of 22 May 1963, blowing from the west-northwest at a speed of 6.26 m/s (14 mph) with higher gusts.\cite{weatherunderground}  This direction points from home plate towards the outfield, which presents us with some difficulty.  Some wind would undoubtedly have topped the stands of Yankee Stadium and impacted the trajectory of the ball as it approached the facade, but, lacking an analysis of how the wind would have flowed around the structural obstacle and its speed as a function of position,  we assume for now that there was no wind.  We will revisit this assumption later, in Section~\ref{sec:plausibility}, but it is a rather significant uncertainty.
    \item Both the magnitude and direction of the ball's spin impact its path.  It is known that certain values of backspin result in a baseball travelling further than simple wind resistance would suggest (unsurprising given the Magnus effect discussed in Section~\ref{sec:theory}), and that this boost has a broad peak at angular velocities $\omega \approx 2200-3000$~rpm before additional backspin may be detrimental to range.\cite{spineffect, nathan_flyball}  We treat this as a variable input to the model and allow for spin to lie between 500 and 3000 rpm, with the spin being entirely backspin so that Equation~\ref{eq:magnus} applies.
    \item The launch angle is another uncertainty, but we can narrow down the range of possibilities via simple considerations.  A straight line between the batter and the facade has an angle of 18.6$^{\circ}$ above horizontal, so the ball likely had a higher launch angle than this; we enforce $\theta_{0} > 10^{\circ}$ to be safe. \footnote{It turns out that Equation~\ref{eq:y_accel} does allow for the Magnus effect to exceed the downward forces of drag and gravity, and for physically plausible speeds.  However, it is implausible for the ball to \textit{maintain} such high speeds through its entire trajectory, so eventually gravity wins out.}  At the upper end of the range of $\theta_{0}$, we rely on the eyewitness reports that the baseball was still rising when it hit the facade; so its launch angle had to be low enough that the trajectory didn't peak prior to the ball's impact.  
\end{itemize}

A given choice of spin and launch angle results in at most one initial speed that matches the boundary values of the home run, so we can plot the trajectory to investigate the minimum speed needed for Mantle's hit to turn out the way it did.

A final thought on the boundary conditions. If we trust eyewitness reports (and this is not a small assumption!) that the ball was still rising when it hit the stadium facade, we may impose an additional constraint on the end of the ball's flight path: it must have $v_{y} > 0$, and not just barely positive.  We assume for simplicity that $v_{y} > 1$~m/s to be the minimum vertical speed that observers could clearly distinguish from a flat or falling trajectory.  This cutoff will serve to eliminate a range of launch angles since some angles would see the ball peak before reaching the target position.

\section{Results}
\label{sec:results}

\begin{figure}[h!]
% The bracketed code determines the figure's placement:  "h" stands for 
% "here", telling LaTeX to put the figure as close to the current location 
% as possible.  The ! overrides LaTeX's tendency to try to find a location 
% that it thinks is better.  But don't agonize over the exact figure placement 
% in your submitted manuscript.  For your initial submission, just make sure 
% each figure is reasonably close to where it's first referenced.
\centering
\includegraphics[width=0.6\textwidth]{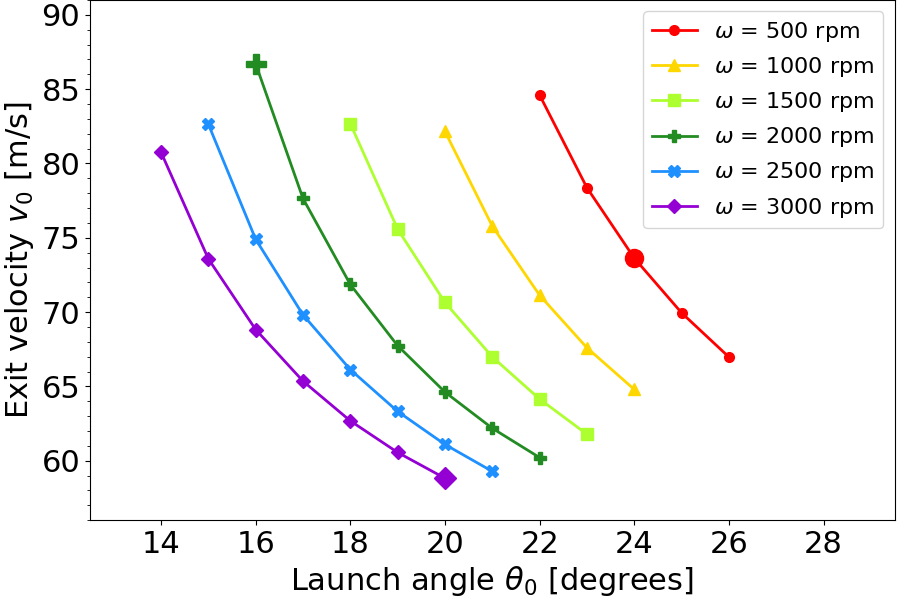}
\caption{Solutions to the boundary-value problem defined in Section~\ref{sec:conditions}.  Each cuve  represents a different amount of backspin $\omega$, and solutions are plotted at integer values (in degrees) for the launch angle $\theta_{0}$.  An upper limit of $v_{0} = 90$~m/s is enforced.  Three specific trajectories are identified with larger symbols; these are the paths shown in Figure~\ref{fig:selected_paths}.}
\label{fig:allowed_speeds}
\end{figure}

Figure~\ref{fig:allowed_speeds} shows a sample set of allowed ``exit velocities''---the common term for the initial speed of the ball as it leaves the bat---that match not only the final position ($x_{f}=106.9$~m, $y_{f}=35.0$~m) but also the final velocity ($v_{y,f} > 1$~m/s) of the baseball.  We considered backspins $\omega$ that were multiples of 500~rpm for simplicity, and likewise restricted launch angle $\theta_{0}$ to integer degree values.  

The curves in the figure are artificially limited to initial speeds of 90~m/s (201 mph) or lower, a speed that is extremely implausible for a human-launched baseball.  The dependence on both launch angle and backspin is clear.  As backspin increases, so does the Magnus effect, meaning the ball  gets more lift and thereby requires a smaller speed to meet our boundary conditions.  Similarly, as launch angle increases, a lower exit velocity is needed to bounce a baseball off of the stadium facade.  The curves terminate at the bottom right of the figure due to our requirement that the ball still be moving upward when it reaches the final position.  Lower initial speeds are physically possible for larger launch angles, but would have resulted in balls that were falling into the facade rather than rising (see also Section VI).

Based on Figure~\ref{fig:allowed_speeds}, it appears that the slowest exit velocity that meets all of our conditions is about 59~m/s (131 mph).  We pick this path and two others for additional examination.

\begin{figure}[h!]
% The bracketed code determines the figure's placement:  "h" stands for 
% "here", telling LaTeX to put the figure as close to the current location 
% as possible.  The ! overrides LaTeX's tendency to try to find a location 
% that it thinks is better.  But don't agonize over the exact figure placement 
% in your submitted manuscript.  For your initial submission, just make sure 
% each figure is reasonably close to where it's first referenced.
\centering
\includegraphics[width=0.6\textwidth]{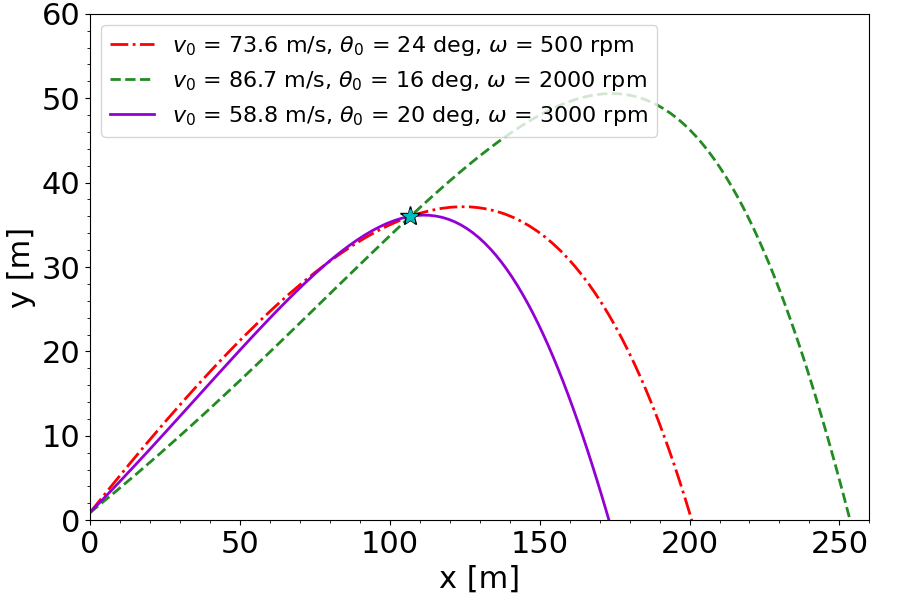}
\caption{Full trajectories for the three sets of initial conditions identified in Figure~\ref{fig:allowed_speeds}. The star in the center of the image is the known final position of the baseball, through which all solutions must pass.  The solution with the lowest initial speed is plotted with a solid purple line; the solution with the longest overall range is plotted with a dashed purple line; and a solution with very low backspin is plotted with a red dash-dotted line.}
\label{fig:selected_paths}
\end{figure}

Our three selected paths are shown in Figure~\ref{fig:selected_paths} and summarized in Table~\ref{tab:results}.  These include the path from Figure~\ref{fig:allowed_speeds} with the slowest exit velocity, the path with the longest range, and then a path with very low backspin for comparison.  The upward arc of the longest path is apparent in the figure for $x\lesssim 50$~m, confirming that the Magnus effect can overcome both drag and gravity.  Comparing the low-spin solution to the the lowest exit velocity, we see that even the slowest initial speed (and also the shortest overall distance) still gets a small amount of lift from the backspin before gravity and wind resistance win out.

\begin{table}[h!]
\centering
\caption{Selected trajectories \& ranges}
\begin{tabular}{l c c r}
\toprule
% The codes above determine the horizontal alignment in each column.
% Options are l (left), r (right), c (centered), and p (paragraph).
% The p option allows an entry to be broken into multiple lines, and
% therefore requires a width specification.
$\omega$ (rpm) & $\theta_{0}$ ($^{\circ}$) & $v_{0}$~(m/s) & Range (m) \\
\midrule
500 & 24 & 73.6 & 201 \\
2000 & 16 & 86.7 & 254 \\
3000 & 20 & 58.8 & 173 \\
\botrule
\end{tabular}
\label{tab:results}
\end{table}
% Tables, like figure captions, should be moved to the end when you submit 
% an editable manuscript for production, after conditional acceptance.
% Put the tables after the endnotes but before the figure captions.

In short, for Mantle's home run, the associated physics and the weather conditions on the day of the event lead to a minimum possible exit velocity of 58.8~m/s (131.5 miles per hour) and a range of 173 meters (568 feet).  These are so far beyond anything tracked with modern technology that we can say with certainty that something is wrong.

In the Statcast era (since 2015) the highest measured exit velocity on a home run was 54.4~m/s (121.7 miles per hour), achieved by Giancarlo Stanton.\cite{hardestHR}  As for the longest home run, the Guinness Book of World Records credits Babe Ruth in 1921 with a range of 175 meters (575 feet), but records indicate that Ruth was batting with an 8.9 m/s (20 mph) tailwind that contributed a great deal to the carry.\cite{guinnessHR} The record in a dome, without wind assistance, may belong to Jose Canseco in 1989, but estimates for the length of his home run range from 480 feet to 540 feet,\cite{longestHR,89ACLS} so it is too unreliable to formally consider.  In the Statcast Era, Nomar Mazara's 153.9-meter (505-foot) home run holds the record.\cite{statcast_longestHR}

At this point we have to concede that either (1) our model from Section~\ref{sec:theory} is wrong, (2) the environmental parameters we used are not right, or (3) the boundary values we imposed are incorrect.  We consider each of these in turn in the following section.

\section{Discussion and plausibility}
\label{sec:plausibility}

%The desired outcome for any activity based on this article is for students to determine a set of initial conditions that allow for all aspects of the home run to be met.  Students (or instructors) who are fans of baseball may wish to go a bit deeper, to determine the most reasonable set of initial conditions and to compare these to known hard-hit balls of the modern era.

Where might our model and assumptions be going astray?

\begin{figure}[h!]
% The bracketed code determines the figure's placement:  "h" stands for 
% "here", telling LaTeX to put the figure as close to the current location 
% as possible.  The ! overrides LaTeX's tendency to try to find a location 
% that it thinks is better.  But don't agonize over the exact figure placement 
% in your submitted manuscript.  For your initial submission, just make sure 
% each figure is reasonably close to where it's first referenced.
\centering
\includegraphics[width=0.4\textwidth]{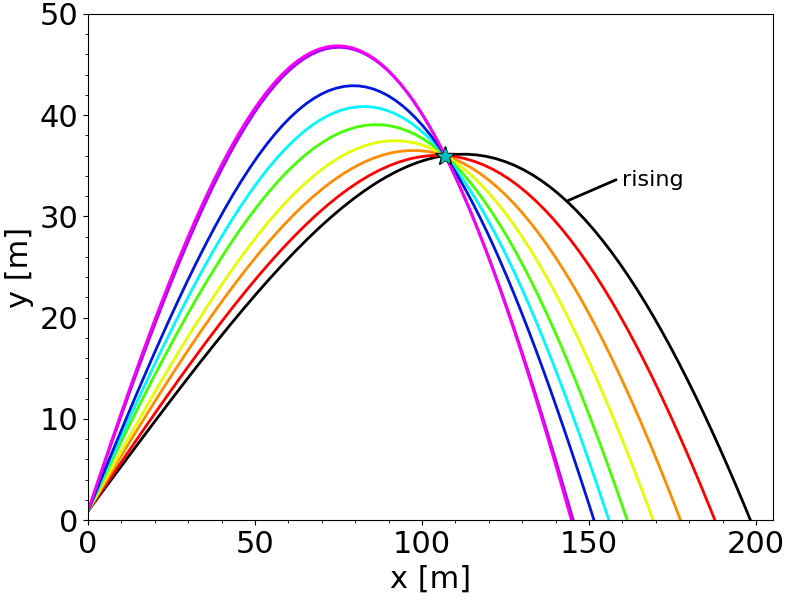}
\includegraphics[width=0.4\textwidth]{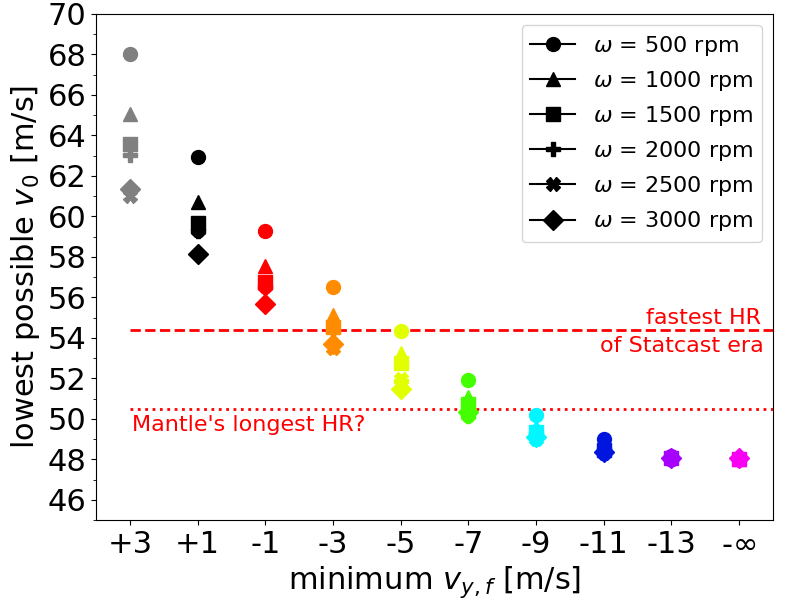}
\caption{\textit{Left panel}: Full trajectories for the various sets of initial conditions described in Section~\ref{sec:plausibility}.  The solid black line is the with-wind trajectory whose exit velocity is 58.1~m/s.  The other eight trajectories are those with the slowest possible initial speed for a given constraint on the final $y$ velocity, also with wind included in calculations.  \textit{Right panel}: the relationship between initial speed and the boundary condition $v_{y,f}$.  Colored dots in here correspond to curves of the same color in the left panel.  Also shown is the exit velocity of the hardest-hit home run since accurate tracking began in 2015.}
\label{fig:other_paths}
\end{figure}

\begin{enumerate}
    \item We will not seriously consider that our model is wrong. The physics presented in Section~\ref{sec:theory} has been validated in countless experiments. 
    \item The largest uncertainties in Section~\ref{sec:conditions} were the presence/strength of wind, the amount of spin, and its direction relative to the ball's velocity vector. Let us consider each of these.
    \begin{enumerate}
       \item As mentioned previously, there was a tailwind on the night of Mantle's home run.  We initially ignored it because it was unclear how to simulate the wind's speed as a function of $x$ and $y$ position, but we reconsider that assumption here.  Adding a constant tailwind (in the $+\hat{x}$ direction), even with a gust to 10~m/s, only slightly reduces the required initial speed, to 58.1~m/s (130~mph).  The tailwind boosts the range of the hit to a whopping 198.3 m (651 feet), which arguably pushes us further away from plausibility.  The true curve for speed vs position likely lies somewhere in between these two extremes, so merely turning on/off wind speed does not resolve the problem.
       \item Lowering the spin of the baseball increases the minimum exit velocity (Figure~\ref{fig:allowed_speeds}), so that is not the solution.  We might try to increase the spin, but such high rates of spin are difficult to reach in conjunction with the high speeds and low launch angles we require: glancing collisions that can impart high spins are more likely to produce popups than home runs.  Indeed, there is evidence that spins above 3000 rpm limit range rather than aiding it.\cite{spineffect, nathan_flyball}  (Spin decay is likely to occur, but is unlikely to be physically significant.\cite{nathan_spindown}  In any case, reducing the spin of the baseball would increase the required exit velocity as already demonstrated.)
       \item We assumed the spin of the baseball was entirely backspin, which is admittedly unlikely.  Adding spin along other axes will not lower the necessary exit velocity; in fact, quite the opposite.  The Magnus effect would then direct the velocity vector of the ball sideways, meaning that a greater initial speed is required so that the outward component of velocity still carries the ball into the facade.
    \end{enumerate}
    It seems, then, that our environmental parameters are already as favorable as we can make them.\footnote{The observed impact of wind on trajectory is affected by the coarse grid of spin rates and launch angles we used.  If we are unlucky with the location of grid points, incorporating wind might result in an \textit{increase} of minimum exit velocity despite intuition suggesting otherwise.  This is not a physical effect, but a numerical one, and discussing this point may be a worthwhile extension of the activity.}
    \item If we adjust the boundary values instead, we can get more reasonable-sounding exit velocities.  It is a known psychological effect that observers are quite bad at determining whether a baseball is rising, especially if they have an ``approaching view'' of the trajectory (i.e. such that the baseball appears to have only vertical motion) rather than a side view.\cite{ShafferMcBeath2005}  The record does not note the seat location of the (unknown) eyewitnesses who claimed the ball was still rising when it struck the stadium, but it is possible that they were in line with the ball's trajectory.  Relaxing the final $y$ velocity of the baseball so that it has passed its apex results in lower exit velocities. Figure~\ref{fig:other_paths} shows the effect of a lower $v_{y,f}$ on both the exit velocity and the trajectory of the baseball; all of the paths in the figure include the tailwind used above.  Even for mildly falling baseballs that just barely passed the peak of their trajectory    (the most likely to be confused with a still-rising path) Mantle's home run exceeds anything achieved in the modern era.  If the ball was falling at up to 3~m/s (i.e. $v_{y,f} > -3$~m/s) when it made contact with the facade, then the minimum necessary exit velocity drops to 53.5~m/s (119.7~mph), a speed that has been eclipsed by modern players.  The range in this case, 177.4 m (582 feet), would still shatter the modern record, so we must again consider our initial conditions.  \footnote{This hypothetical home run would have easily exceeded Mantle's longest hit    for which there is compelling historical evidence.  Analysis\cite{mantle_longball} of his home run on 17 April 1953 yields an estimated distance of 164 m (538 feet), but an exit velocity of ``only'' 50.5 m/s (113 mph).}  A steeper initial angle results in still-lower initial speeds, at the cost of being less likely that observers would have reported the baseball as still rising.
\end{enumerate}

Recall now that we chose the most favorable possible spin conditions, so the true exit velocity of the ball would be slightly higher to make up for the less-than-perfect spin.  The speeds in Figure~\ref{fig:other_paths} are consequently likely to be underestimates, but to what degree we cannot be certain.  However, almost all trajectories in Figure~\ref{fig:other_paths} exceed 154 meters in range, which would make them the longest home run in the modern era even if they weren't the fastest.

\section{Conclusions}
\label{sec:conclusions}

In short, our simple model for examining the flight of Mickey Mantle's self-professed hardest-hit home run leads to unbelievable conclusions.  There does not appear to be a single set of initial conditions that satisfies (1) known atmospheric conditions, (2) eyewitness reports, and (3) the plausible limits of human effort.  If we assume the eyewitnesses were mistaken about the rising trajectory of the baseball, then Mantle's home run no longer requires superhuman strength.  Instead, it trades record-breaking initial speed for record-breaking distance, bringing us back to the idea presented in the introduction.  Barring breakthroughs in cloning or time travel, we won't ever get to directly pit legends of baseball against modern-day superstars.  But we can apply the rules of physics to get a glimpse of just how special those legends were---and why the sport continues to hold the attention and imagination of fans around the world.

\backmatter

\bmhead{Acknowledgments}

We gratefully acknowledge Manasvi Lingam for the initial push to turn a classroom activity into the present manuscript, and Alan Nathan for extremely helpful discussions on the spin and drag coefficients of batted baseballs.  We would also like to thank the anonymous referees whose numerous comments, including the psychology of observing fly balls, improved the physics and the presentation of the manuscript.

\bmhead{Author declarations}

The authors have no conflicts to disclose.

\end{document}